\documentclass[]{emulateapj}

\begin{document}


\newcommand{\lsim}{\lower0.6ex\vbox{\hbox{$ \buildrel{\textstyle <}\over{\sim}\ $}}}
\newcommand{\gsim}{\lower0.6ex\vbox{\hbox{$ \buildrel{\textstyle >}\over{\sim}\ $}}}


\newcommand{\hvol}{h^{3}{\mathrm{Mpc}}^{-3}}
\newcommand{\hmpc}{h^{-1}\mathrm{Mpc}}
\newcommand{\hkpc}{h^{-1}\mathrm{kpc}}
\newcommand{\hpc}{h^{-1}\mathrm{pc}}
\newcommand{\hMsun}{\ h^{-1}\mathrm{M}_{\odot}}
\newcommand{\hMpc}{\ h^{-1}\mathrm{Mpc}}
\newcommand{\Msun}{M_{\odot}}
\newcommand{\kms}{{\,{\mathrm{km}}\,{\mathrm{s}}^{-1}}}
\newcommand{\kpc}{{\,{\mathrm{kpc}}}}
\newcommand{\Gyr}{{\,{\mathrm{Gyr}}}}

\newcommand{\rhomean}{\rho_{\mathrm{M}}}

\newcommand{\mpt}{m_{\mathrm{p}}}
\newcommand{\rfind}{r_{\mathrm{f}}}


\newcommand{\Mr}{M_r^h}
\newcommand{\Mvir}{M_{\mathrm{vir}}}
\newcommand{\Rvir}{R_{\mathrm{vir}}}
\newcommand{\Dvir}{\Delta_{\mathrm{vir}}}
\newcommand{\Vmax}{V_{\mathrm{max}}}

\newcommand{\Vmaxa}{V_{\mathrm{max, acc}}}
\newcommand{\Vmaxn}{V_{\mathrm{max, now}}}

\newcommand{\vmax}{V_{\mathrm{max}}}
\newcommand{\vmaxa}{V_{\mathrm{max, acc}}}
\newcommand{\vmaxn}{V_{\mathrm{max, now}}}

\newcommand{\Vhost}{V_{\mathrm{host}}}
\newcommand{\Vsat}{V_{\mathrm{sat}}}

\newcommand{\Nsat}{N_{\mathrm{sat}}}
\newcommand{\Ngal}{N_{\mathrm{gal}}}

\newcommand{\LCDM}{$\Lambda$CDM}

\shortauthors{Mar\'in, Wechsler, Frieman and Nichol}
\shorttitle{Modeling galaxy three-point statistics}

\title{Modeling the galaxy three-point correlation function}
\author{Felipe A. Mar\'in\altaffilmark{1,5}, 
  Risa H. Wechsler\altaffilmark{1,2}, Joshua A. Frieman\altaffilmark{1,3} and  Robert C. Nichol\altaffilmark{4} 
}
\altaffiltext{1}{
  Department of Astronomy \& Astrophysics,
  Kavli Institute for Cosmological Physics,
  The University of Chicago, 
  Chicago, IL 60637 USA
}
\altaffiltext{2}{
  Kavli Institute for Particle Astrophysics \& Cosmology,
  Physics Department, and Stanford Linear Accelerator Center,
  Stanford University,
  Stanford, CA 94305
}
\altaffiltext{3}{
  Center for Particle Astrophysics, Fermi National Accelerator Laboratory, 
  P.O. Box 500, Batavia, IL 60510 USA
}
\altaffiltext{4}{Institute of Cosmology \& Gravitation, University of Portsmouth,
Portsmouth, PO1 2EG, UK}

\altaffiltext{5}{
e-mail: fmarinp@uchicago.edu
}

\begin{abstract}
  We present new predictions for the galaxy three-point
  correlation function (3PCF) using
  high-resolution dissipationless cosmological simulations of a flat
  $\Lambda$CDM Universe which resolve galaxy-size halos and
  subhalos. We create realistic
  mock galaxy catalogs by assigning luminosities and colors to dark
  matter halos and subhalos, and we measure the reduced 3PCF as a
  function of luminosity and color in both real and redshift 
  space. 
  As galaxy luminosity and
  color are varied, we find small differences in the amplitude and
  shape dependence of the reduced 3PCF, at a level qualitatively
  consistent with recent measurements from the SDSS and 2dFGRS. 
  We confirm that discrepancies between previous 3PCF measurements can
  be explained in part by differences in binning choices. We explore 
  the degree to which a simple local bias model can fit the 
  simulated 3PCF. The
  agreement between the model predictions and galaxy 3PCF measurements
  lends further credence to the straightforward association of
  galaxies with CDM halos and subhalos.

\end{abstract}

\maketitle

\section{Introduction}

Observations of the higher-order statistics of the galaxy distribution
can provide fundamental tests of the standard cosmological model. For
example, higher-order correlation functions of the mass are
predicted to be zero in linear perturbation theory for Gaussian
initial conditions, which are expected in the simplest inflation
models of the early Universe (\citealt{peebles:80},
\citealt{bernardeau_etal:02}, \citealt{szapudi:05} and references
therein). In the late Universe, however, non-linear gravitational
clustering and biased galaxy formation lead to non-Gaussianity in the
galaxy density field, resulting in non-zero connected N-point
correlation functions (NPCFs) with $N>2$. By studying higher-order
galaxy statistics on large scales, we can test the nature of the
initial conditions; on smaller scales, the NPCFs can constrain models
of biased galaxy formation (e.g., \citealt{fry_gaztanaga:93,
frieman_gaztanaga:94}) and the relationship between galaxies and their
host dark matter halos.

The three-point correlation function (3PCF, or $\zeta$), and its
Fourier-space equivalent, the bispectrum, are the first in the
hierarchy of higher-order statistics and measure the shape dependence
of the number of galaxy triplets as a function of scale. The 3PCF is
sensitive to, for example, the shapes of dark matter halos and the presence of 
filamentary structures in the large-scale structure of the
Universe (\citealt{sefusatti_scoccimarro:05}). Since the pioneering
work of Peebles and collaborators (see \citealt{peebles:80} and
references therein, in particular \citealt{groth_peebles:77}), there have been many measurements of the 3PCF and
bispectrum using a variety of angular
\citep{peebles_groth:75,groth_peebles:77,gaztanaga_frieman:94,
  frieman_gaztanaga:99,huterer_etal:01,szapudi_etal:01,szapudi_etal:02,ross_etal:06}
and redshift \citep{gaztanaga_frieman:94,jing_borner:98,verde_etal:98,
scoccimarro_feldman:01,feldman_etal:01,jing_borner:04} catalogs. 

There has also been
considerable theoretical work to understand the 3PCF and bispectrum
using non-linear perturbation theory (see \citealt{bernardeau_etal:02} and
references therein), the halo model (see \citealt{ma_fry:00,scoccimarro_etal:01,wang_etal:04,fosalba_etal:05}),
 and cosmological simulations
\citep{barriga_gaztanaga:02,scoccimarro_etal:99,gaztanaga_scoccimarro:05,hou_etal:06}.

In recent years, there has been renewed interest in the 3PCF due to
the availability of large redshift surveys.
These surveys now provide both the volume and the number of
galaxies required to make robust measurements 
of the 3PCF over a range of scales. For
example, recent papers by \citet{kayo_etal:04},
\citet{hikage_etal:05}, \citet{nichol_etal:06}, \citet{nishimichi_etal:06}, 
\citet{kulkarni_etal:07} have presented
measurements of the 3PCF from the Sloan
Digital Sky Survey (SDSS; \citealt{york_etal:00}) as a function of
scale, galaxy luminosity, and color.  \cite{nichol_etal:06} also
quantified the effect of large-scale structures on the shape
dependence of the 3PCF. Likewise, several recent papers
\citep{jing_borner:04,baugh_etal:04,
  croton_etal:04,gaztanaga_etal:05,pan_szapudi:05,croton_etal:06b} provide new
measurements of the 3PCF and high-order correlations from 
the Two-degree Field Galaxy
Redshift Survey (2dFGRS; \citealt{colless_etal:01}).

The main results from these recent SDSS and 2dFGRS
analyses of the 3PCF are: {\it i)} on large scales, the observed 3PCF
is in qualitative agreement with expectations for the growth of
structure from Gaussian initial conditions; {\it ii)} on smaller
scales (i.e., in the non-linear and weakly non-linear regimes), the
3PCF measured in redshift space scales with the redshift-space
two-point correlation function (2PCF, or $\xi$) as $\zeta\sim\xi^2$,
consistent with the ``hierarchical clustering'' ansatz 
\citep{peebles:80}; {\it iii)} the shape dependence of the reduced 3PCF 
depends at most weakly on galaxy luminosity; 
{\it iv)} the amplitude of the 3PCF is larger for elongated
triangle configurations than for more symmetric triangle shapes
\citep{gaztanaga_scoccimarro:05,gaztanaga_etal:05,nichol_etal:06, 
kulkarni_etal:07} --- again consistent with expectations from 
non-linear clustering theory.

It is important to make detailed comparisons of these 3PCF
observations with theoretical predictions that incorporate a realistic 
prescription for modeling galaxies and that account for observational 
effects such as redshift-space distortions.  
Such comparisons can be carried out in two ways: either the observations
can be corrected for the redshift-space effects and compared to theory
in real space, e.g., using the projected 3PCF \citep{zheng:04}, which
is analogous to the projected 2PCF
\citep{zehavi_etal:05}, or one can build mock galaxy
catalogs from cosmological simulations and measure the theoretical
3PCF directly in redshift space. Since we also have access to
the real-space 3PCF from such mock catalogs, we can
investigate in detail the relationship between the real- and
redshift-space correlation functions.

In this paper, we pursue the second of these methodologies, using
state-of-the-art high-resolution dissipationless dark matter
cosmological simulations.  These simulations have the spatial resolution
required to identify the dark matter (DM) halos and subhalos that host
individual galaxies and at the same time encompass a large enough 
volume to probe large-scale structure in a statistically reliable way.
The model we use assigns galaxy properties (luminosity, color,
etc.) directly to these DM galactic halos and subhalos using simple, 
empirically-based assumptions about these properties.  This approach differs in
both assumptions and the resolution required from methods which
build galaxy catalogs by statistically assigning several galaxies to
each (more massive) halo using a Halo Occupation Distribution (HOD;
e.g.  \citealt{berlind_weinberg:02,wang_etal:04}; see \citet{kulkarni_etal:07} for
constraints on HOD parameters from the SDSS Luminous Red Galaxy sample's 3PCF) or from semi-analytic 
methods. The method used here was first
implemented by \cite{kravtsov_etal:04}, and has been applied
successfully to different statistical studies, including the 2PCF
\citep{conroy_etal:06}, galaxy-galaxy lensing
\citep{tasitsiomi_etal:04}, and close pair statistics
\citep{berrier_etal:06} among others \citep[see
also][]{vale_ostriker:04, vale_ostriker:06, conroy_etal:07}.  Here we
extend the study of this model to the 3PCF as a function of
luminosity, color, and redshift.  Where possible, we make direct
comparisons in redshift space with recent observations.

In \S \ref{sec:sim}, we describe the simulations used in this paper and our
methods for constructing mock galaxy catalogs based on resolved DM
halos. We also review the techniques used to estimate the NPCFs.  In
\S \ref{sec:meas}, we present measurements of the 3PCF in 
both real and redshift
space for both the dark matter and galaxy catalogs, 
while in \S \ref{sec:lumcolor}, we
study the dependence of the model 3PCF on galaxy
luminosity and color. In \S \ref{sec:compare}, we compare the model 3PCF 
with SDSS observations and discuss the effects of binning. We 
relate the 3PCF to a simple non-linear bias model in \S \ref{sec:bias}. 
In \S \ref{sec:conclude}, we summarize and
discuss our findings.

\section{Methods}
\label{sec:sim}
\subsection{Dark matter simulations}

We investigate clustering statistics using cosmological $N$-body
simulations of structure formation in the concordance, flat
$\Lambda$CDM cosmology with $\Omega_{\Lambda}=0.7=1-\Omega_m$,
$h=0.7$, and $\sigma_8=0.9$, where  $\Omega_m$, $\Omega_\Lambda$ are
the present matter and vacuum densities in units of the critical
density, $h$ is the Hubble parameter in units of 100 km s$^{-1}$
Mpc$^{-1}$, and $\sigma_8$ specifies the present linear rms mass fluctuation in
spheres of radius 8 $h^{-1}$Mpc.  The simulations used here
were run using the Adaptive Refinement Tree $N-$body code (ART, see
\citealt{kravtsov_etal:97} for details), which implements successive
refinements in space and time in high-density environments.  The
primary simulation box we use is 120 $h^{-1}$Mpc on a side (hereafter,
L120); the number and mass of each dark matter particle are
$N_p=512^3\approx1.34\times 10^8$ and $m_p=1.07\times 10^9
h^{-1}M_{\odot}$ respectively.  This simulation has been previously
used to measure several properties of dark matter halos and subhalos
\citep[e.g.,][]{allgood_etal:06,wechsler_etal:06,conroy_etal:06,
  berrier_etal:06}.  In order to include more massive halos and study
the effects of the size of the sample on the statistical analysis, we
also use a second simulation with the same cosmological parameters in
a bigger box, with 200 $h^{-1}$Mpc on a side \cite[which was also used
to measure halo shapes in][]{allgood_etal:06}. This box contains
$N_p=256^3$ particles with mass $m_p=3.98\times 10^{10}$ $h^{-1}M_{\odot}$,
therefore it will lack low mass (and luminosity) objects that are
included in the L120 box.

From these dark matter samples, virialized concentrations of particles
are identified as halos. In order to find these halos and their
constituent subhalos (concentrations of virialized matter inside
bigger halos), a variant of the Bound Density Maxima halo finding
algorithm of \cite{klypin_etal:99} is used.  This algorithm assigns
densities to each particle using a smoothing kernel on the 32 nearest
neighbors; centering on the highest-overdensity particle, each center
is surrounded by a sphere of radius $r_{\rm find}=50h^{-1}$kpc.
The algorithm removes unbound particles when calculating the
properties of the halos. The halo catalog is complete for halos with more
than 50 particles, which corresponds to a minimum halo mass of 
$1.6\times 10^{10}$  $h^{-1}M_{\odot}$ for the L120 box and $2.0\times10^{12}$
 $h^{-1}M_{\odot}$ for halos in the L200 box.

Henceforth, we will use the terms ``distinct halo'' to mean any halo
that is not within the virial radius of a larger halo, ``subhalo'' to
indicate a halo that is within the virial radius of a larger halo, and
``galactic halo'' to refer to the halo directly hosting a galaxy.
Using this terminology, the galactic halo of a satellite galaxy is a
subhalo while the galactic halo of a central or isolated galaxy will
be a distinct halo.

\subsection{From halos to `galaxies'}
It is expected that galaxy properties depend in detail not only on the dark
matter clustering, but also on the gas dynamics, radiative 
processes, and feedback mechanisms that 
affect the baryonic components. A program to include all those 
physical processes in building mock galaxy catalogs from simulations
would require introducing a number of free parameters and assumptions which 
could partially obscure the relevant mechanisms that determine the 
shape and amplitude of the 3PCF. Our approach is instead more 
empirical---to associate galaxies of given properties with simulated 
dark matter halos and subhalos, using halo properties that 
we can measure in the simulation, and to see whether this one-to-one 
correspondence predicts a galaxy 3PCF that is 
consistent with the observations. As described below, the assignment 
of galaxies to halos was designed to reproduce certain  
features of the observed galaxy distribution, but the 3PCF was not 
one of these. As a result, the 3PCF constitutes a non-trivial 
test of this approach.

The primary galaxy samples used here are created by assigning galaxy
luminosities and colors drawn from the SDSS redshift survey to dark
matter halos and subhalos, using the maximum circular velocity at
$z=0$, $\vmax$, as an indicator of the halo virial mass. $\vmax$ has 
been found to be a good proxy of the galaxy potential well,
which is a good indicator of stellar mass \citep{kravtsov_etal:04,
conroy_etal:06}.

A galaxy luminosity
in the $r-$band is assigned to each halo by matching the cumulative
velocity function $n(>\vmax)$ of all galactic halos (distinct halos
and subhalos) to the observed SDSS $r$-band luminosity function
\citep{blanton_etal:03} at $z=0.1$ (the approximate mean redshift of
the main spectroscopic SDSS galaxy sample used to estimate the
luminosity function).  To correct to $z=0$ magnitudes, we use the code
\texttt{kcorrect v3\_2} \citep{blanton_etal:03kcorr}.  Since the
limited size of the box gives us an upper limit on the luminosities
which can be reliably studied (in the statistical sense), and at the
same time we cannot sample the lowest-luminosity objects due to
limited spatial resolution, for the L120 box we present results for
galaxies in the absolute magnitude range $-19\geq \Mr \geq -22$, where
$\Mr \equiv M_r-5\log h$.  The L200 box has a lower spatial resolution,
therefore it contains only brighter objects, with $\Mr \lsim -20$.  In
order to assign colors to the galaxies, we use the procedure described
by \citet{wechsler:04} and \citet{tasitsiomi_etal:04}. This method
uses the relation between local galaxy density (defined as the
distance to the tenth nearest neighbor brighter than $\Mr = -19.7$ and
within $cz=1000$ km/s) and color observed in the SDSS, using the
CMU-Pitt Value Added Catalog constructed from DR1
\citep{abazajian_etal:03}, to assign a color to each mock galaxy.  We
use the distant observer approximation \citep{bernardeau_etal:02} to
obtain positions in redshift space.  Table \ref{tab:samples} describes
the subsamples used in this study.

Although the general association of galaxies with dark matter subhalos
seems quite robust, the detailed association of galaxy properties with
subhalo properties is less clear.  While galaxy luminosity is expected
to be quite tightly connected to velocity, the circular velocity of a
given subhalo decreases with time due to tidal stripping as it
interacts with its host halo.  Observable galaxy properties such as luminosity and color 
will likely be less
affected by this process.  This implies that galaxy observables may be
more strongly correlated with $\vmaxa$, the maximum circular velocity 
of the halos at the moment of accretion onto their host, than with 
the current maximum circular velocity, $\vmaxn$. This conclusion is 
supported by measurements of two-point statistics on both large and
small scales in simulations 
\citep{conroy_etal:06, berrier_etal:06}.  Motivated by
these considerations, we construct galaxy catalogs using both 
 $\vmaxa$ and $\vmaxn$.  We also use
galaxy catalogs at different redshifts in order to study the
evolution of the 3PCF with time. All these additional catalogs use the
L120 box and have the same spatial density as the sample of halos
selected by $\vmaxn$.  We note that the model for assigning color is
also not uniquely determined.  It is, however, sufficient to match the
two-point clustering length for red and blue galaxies and several observed 
properties of galaxy clusters \citep{zehavi_etal:05}.  
Future measurements of both clustering
statistics and of the properties of galaxies in groups and clusters
should help to further refine these galaxy assignment models.
%
\begin{deluxetable}{ l l c c}
\tablecolumns{4}
\tablecaption{Subsamples at $z=0$}
\tablehead{
\colhead{Box} & \colhead{Subsample} & \colhead{Number} & \colhead{Density}\\
\colhead{} & \colhead{} & \colhead{of objects}& \colhead{[($h^{-1}$Mpc)$^{-3}$]}}
\startdata
L120 & All objects $\Mr<-19$  & 25371 &$1.5\times10^{-2}$ \\
L120 & $-19<\Mr<-20$ & 15432 & $8.9\times10^{-3}$ \\
L120 & $-20<\Mr<-21$ & 8153  & $4.7\times10^{-3}$\\
L120 & red ($g-r>0.7$) & 10059 &$5.8\times10^{-3}$  \\
L120 & blue ($g-r<0.7$) & 15312 & $8.9\times10^{-3}$\\
L200 & All objects $\Mr<-20$  & 43564 & $5.4\times10^{-3}$\\
L200 & $-20<\Mr<-21$ & 30575 & $3.8\times10^{-3}$ \\
L200 & $-21<\Mr<-22$ & 9921  & $1.2\times10^{-3}$\\
L200 & red ($g-r>0.7$) & 17278 &$2.2\times10^{-3}$  \\
L200 & blue ($g-r<0.7$)& 23748 & $2.9\times10^{-3}$\\
\\
\enddata
\label{tab:samples}
\end{deluxetable}

%

\subsection{Measuring the 3PCF}

Just as the 2PCF measures the excess probability of finding two objects
separated by a distance $r$, the 3PCF describes the probability of finding three
objects in a particular triangle configuration compared to a random
sample. The probability of finding three objects in three arbitrary volumes
$dV_1$, $dV_2$, and $dV_3$, at positions $r_1$, $r_2$ and $r_3$ respectively,  
is given by \citep{peebles:80}
\begin{eqnarray}\label{eq:P3}
P &=& [1+\xi(r_{12})+\xi(r_{23})+\xi(r_{31})+\zeta(r_{12},r_{23},r_{31})] \times \nonumber\\ && \bar{n}^3 dV_1 dV_2 dV_3,
\end{eqnarray}
where $\bar{n}$ is the mean density of the objects, $r_{ij}\equiv r_i-r_j$ the distance between 
two objects, $\xi$ is the
2PCF, and $\zeta$ is the 3PCF:
\begin{eqnarray}
\xi(r_{12}) &=& \langle\delta(r_1)\delta(r_2)\rangle \\
\zeta(r_{12},r_{23},r_{31})&=&\langle\delta(r_1)\delta(r_2)\delta(r_3)\rangle;
\end{eqnarray}
here $\delta$ is the fractional overdensity in the dark matter field or
in the distribution of galaxies.
 Since the 3PCF depends
on the configuration of the three distances, it is sensitive to the
2-D shapes of the spatial structures, at large and small scales 
\citep{sefusatti_scoccimarro:05,gaztanaga_scoccimarro:05}.
Motivated by the ``hierarchical'' form of the N-point functions,
$\zeta \propto \xi^2$, found by \cite{peebles_groth:75},  
we use the reduced 3PCF $Q(r,u,\alpha)$ 
to present our results:
\begin{eqnarray}
Q(r,u,&\alpha&)=\nonumber\\ 
\frac{\zeta(r,u,\alpha)}{\xi(r_{12})\xi(r_{23})+ \xi(r_{23})\xi(r_{31})+\xi(r_{31})\xi(r_{12})}.
\end{eqnarray}
This quantity is useful since $Q$ is found to be close to
unity over a large range of scales even though $\xi$ and $\zeta$  
vary by orders of magnitude (\citealt{peebles:80}).
To parametrize the triangles for the 3PCF measurements, 
$r\equiv r_{12}$ sets the scale size of the triangle, while the shape
parameters are given by the ratio of two sides of the triangle, 
$u=r_{23}/r_{12}$, and the angle between the two sides of the triangle
$\alpha=\cos^{-1}(\hat{r}_{12}\cdot \hat{r}_{23})$, 
where $\hat{r}_{12}$, $\hat{r}_{23}$ are the unit vectors of the first
two sides.
Following \cite{gaztanaga_scoccimarro:05}, triangles where
$\alpha$ is close to 0$^\circ$ or 180$^\circ$ are referred to as
``elongated configurations'', while those with $\alpha \sim 50^\circ -
120^\circ$ are referred to as ``rectangular configurations'',

We calculate
the 2PCF using the estimator of \cite{landy_szalay:92}, 
\begin{equation}
\xi = \frac{DD-2DR+RR}{RR}.
\end{equation}
Here, $DD$ is the number of data pairs normalized by $N_D\times
N_{D}/2$, $DR$ is the number of pairs using data and random catalogs
normalized by $N_DN_R$, and $RR$ is the number of 
random data pairs normalized by
$N_R\times N_R/2$.  The 3PCF is calculated using the
\cite{szapudi_szalay:98} estimator:
\begin{equation}
\zeta=\frac{DDD-3DDR+3DRR-RRR}{RRR},
\end{equation}
where $DDD$, the number of data triplets, is normalized by $N_D^3/6$,
and $RRR$, the random data triplets, is normalized by $N_R^3/6$. $DDR$ is
normalized by $N_D^2N_R/2$, and $DRR$ by $N_DN_R^2$.

We estimate the errors using jack-knife re-sampling. From 
each galaxy catalog,  we construct
sixteen subsamples of L120 or L200; within each of them we remove a 
different region ($30\times 60^2$ ($h^{-1}$Mpc)$^3$ for the L120 box, 
and $50\times 100^2$ ($h^{-1}$Mpc)$^3$ for the L200 box).
The variance $\sigma_{JK}$ of $Q$ is calculated as:
\begin{equation}
\sigma_{JK}^2 = \frac{N-1}{N}\sum_{i=1}^N(Q_i-\bar{Q}_i)^2,
\end{equation}
where $N=16$ is the number of subsamples, $Q_i$ is the value for the
$i-$th subsample, and $\bar{Q}_i$ is the mean of the $Q_i$.
We note in passing that the validity of jack-knife resampling as a method to
estimate the errors has not been explicitly tested with mock catalogs for
three-point statistics. Although it is beyond the scope of this paper, this 
may be an interesting topic of future investigation
especially once the statistical power of the measurements improves.

To compute the 2PCF and 3PCF, we use the {\it NPT} software developed
in collaboration with the Auton Lab at Carnegie Mellon
University. {\it NPT} is a fast implementation of the NPCFs using
multi-resolution kd-trees to compute the number of pairs and
triplets in a dataset. For more details and information on the
algorithm, see \cite{moore_etal:01}, \cite{gray_etal:04}, and
\citet{nichol_etal:06}.

\section{The 3PCF of galaxies and dark matter}
\label{sec:meas}

We estimate the reduced 3PCF for the distribution of dark matter
and for galaxies for different triangle configurations, focusing
on the scale and shape dependence of the 3PCF.  We also investigate its time
evolution and how it depends on the selection criterion for
subhalos. We study $Q(r,u,\alpha)$ in both real and 
redshift space, in order to compare our results with current
observations and disentangle galaxy biasing effects from those which
are consequences of redshift distortions.

In order to distinguish scale and, most importantly, shape effects,
and to keep the errors as small as possible, we have chosen an
intermediate-resolution binning scheme. For studies of equilateral
triangles ($u=1$ and $\alpha = \pi/3$ rad), 
we use bins of size $\Delta \log(r) = 0.1$.
For measurements of the shape dependence of the
3PCF, we use triangles with four different scales $r=$ 0.75, 1.5, 3, 6, 
and 9 $h^{-1}$Mpc, using the shape parameters $u = 2$,  and 15 angular bins separated by 
$\Delta \alpha = \pi/15$ rad; the resolution of the bins is given by $\Delta r_{ij} = \pm 0.03r_{ij}$.
 This resolution is sufficient to see the most important
features of the 3PCF even on small scales, although it is not
sufficient to distinguish the ``finger-of-god'' effect at the smallest scales
in redshift space, where $Q(\alpha)$ varies very little except at very 
small or elongated angles, where it increases to many times the mean value
(\citealt{gaztanaga_scoccimarro:05}).

\begin{figure*}
\plottwo{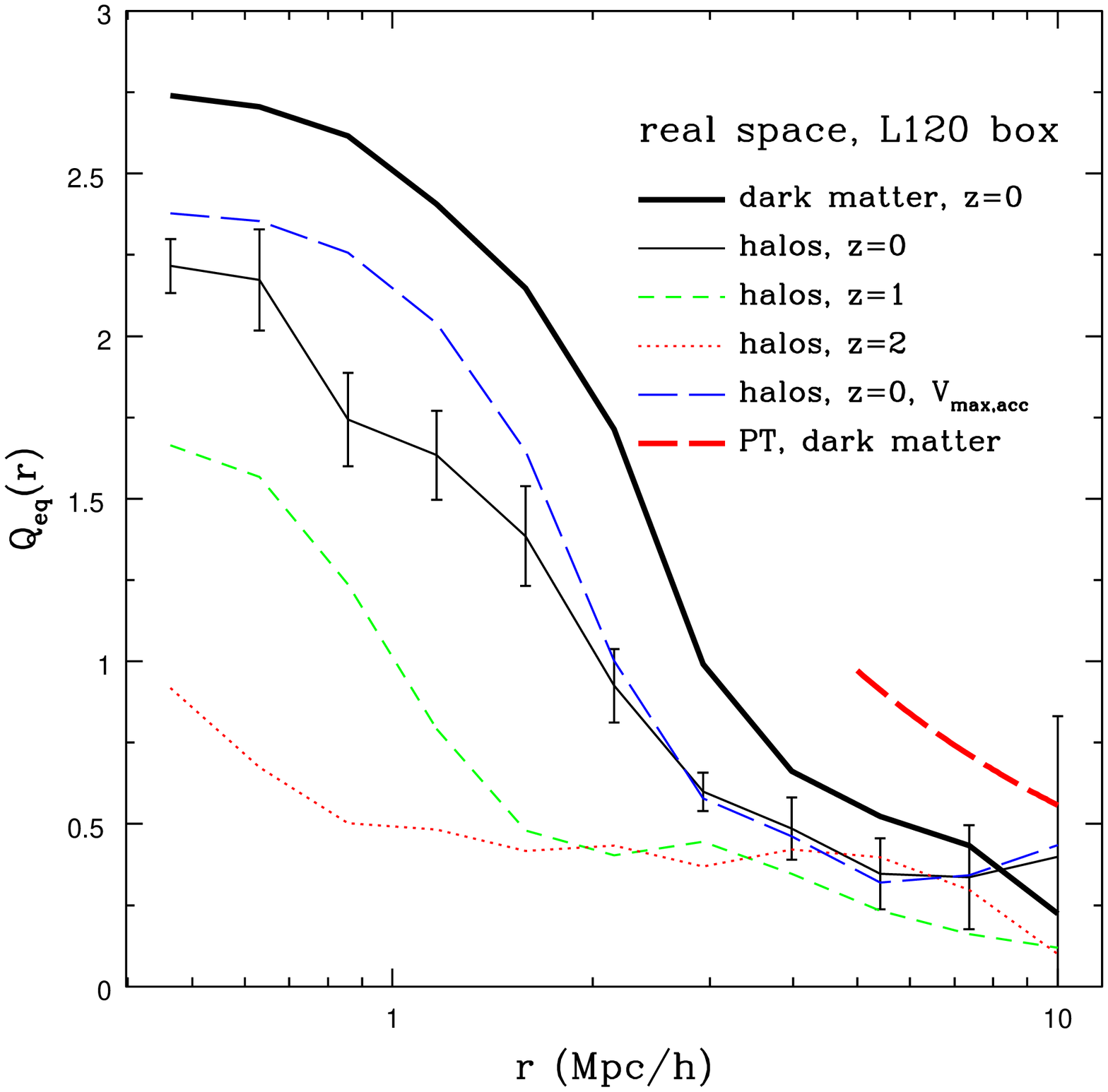}{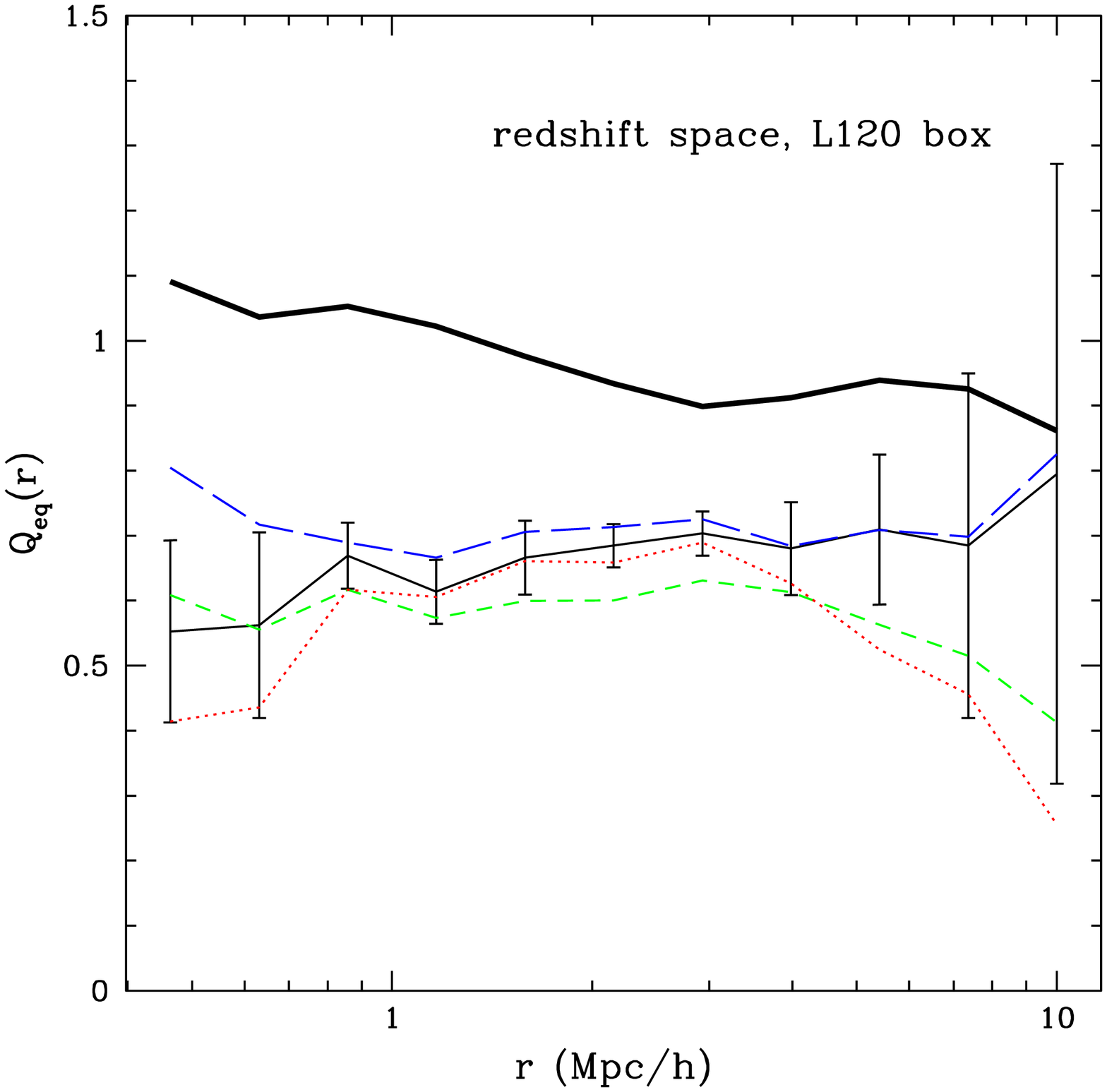}
\caption{The reduced 3PCF, $Q_{eq}(r)$, for equilateral triangles
  in the L120 box. \emph{Left:} Results in real space. \emph{Thick solid line (black):} dark
  matter; \emph{thin solid line (black):} galactic halos selected by $V_{max,now}$
  at $z=0$; \emph{short-dashed (green):} galactic halos selected by
  $V_{max,now}$ at $z=1$; \emph{dotted (red):} galactic halos
  selected by $V_{max,now}$ at $z=2$; \emph{long-dashed (blue):}
  galactic halos selected by $V_{max,acc}$ at $z=0$; \emph{thick long-dashed 
(red):} leading-order perturbation theory, dark matter. \emph{Right:}
  Results in redshift space. Line types
  correspond to the same dark matter and halo samples as in the
  left-hand plot. Error bars are calculated using jack-knife resampling and 
are only shown for one of the samples for clarity.}
\label{fig:equ3PCF} 
\end{figure*}

\subsection{The 3PCF in real space}

\begin{figure*}
\plottwo{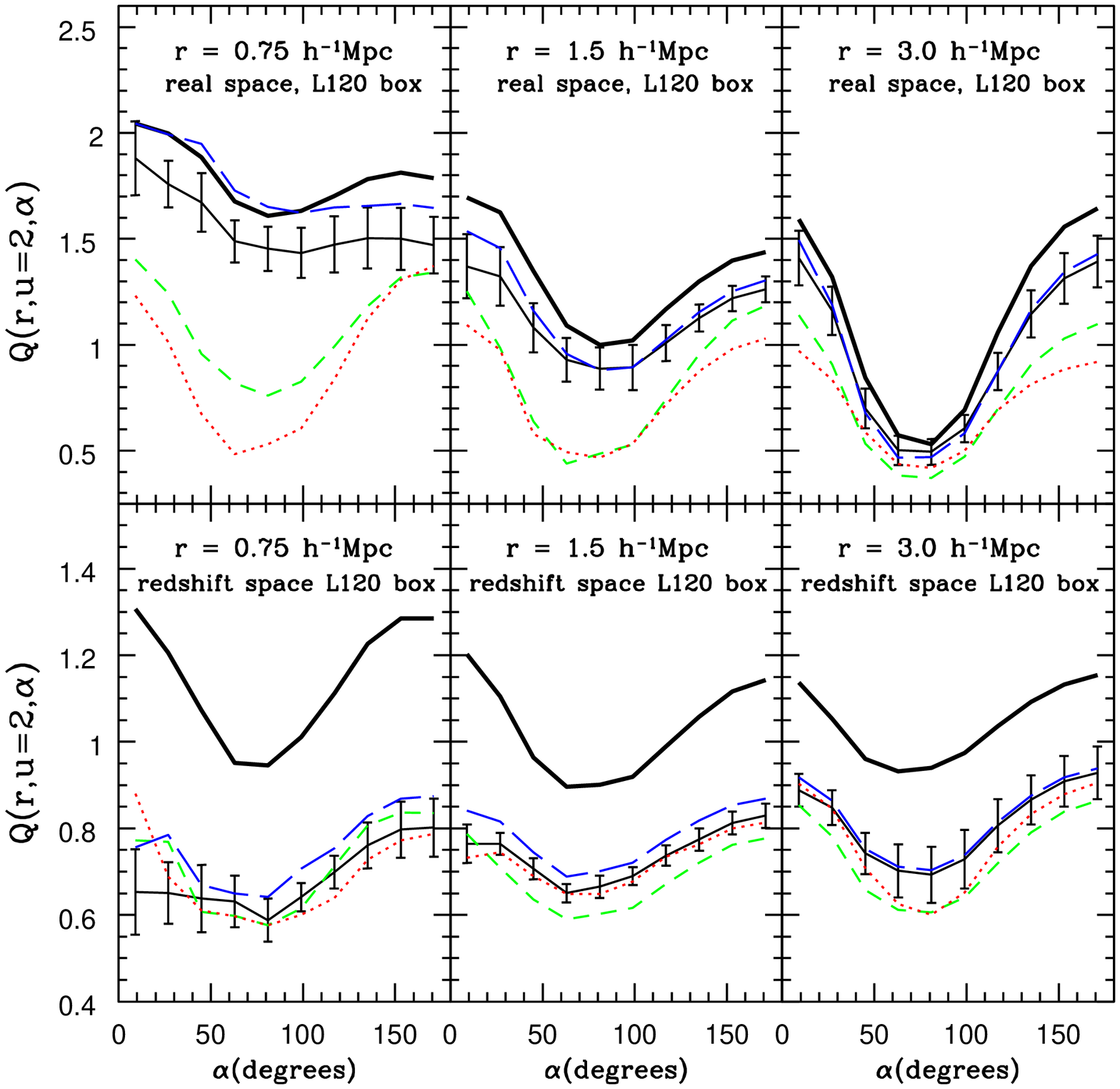}{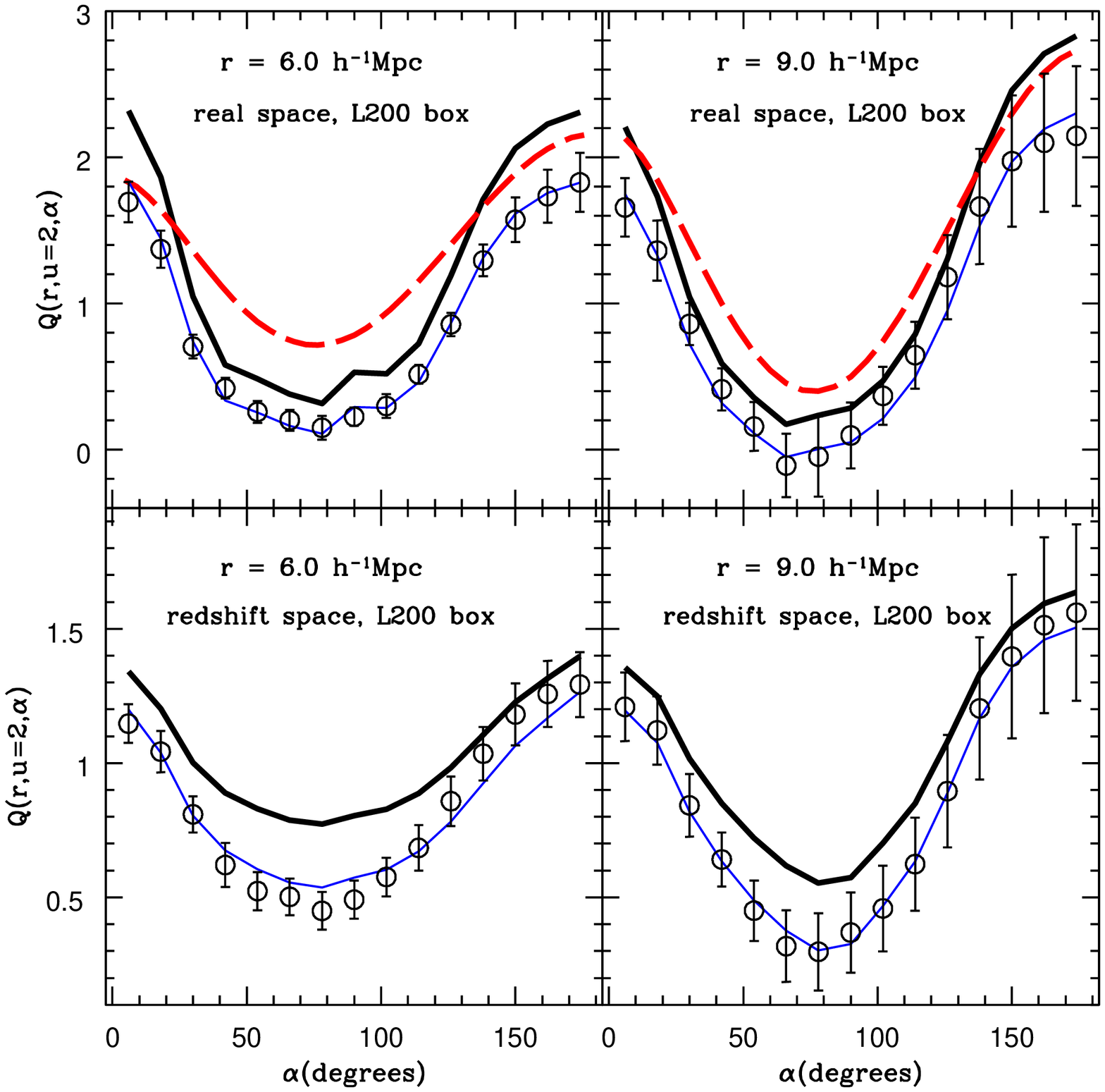}
\caption{ Measurement of the reduced 3PCF as a
  function of triangle shape, $Q(\alpha)$, 
  for different scales $r$, with side ratio $u=r_{23}/r_{12}=2$ fixed.
  \emph{Left:} Results in the L120 box for  $r=0.75, 1.5$, and $3$ $h^{-1}$Mpc
  (from left to right) in real space (top panels) and redshift space (bottom panels) for 
  dark matter and galaxies; line types
  correspond to the same dark matter and halo samples as in Figure 1.
  \emph{Right:} Results in the L200 box for  $r=6$ (left) and $9$ $h^{-1}$Mpc (right)
  in real (top) and redshift space (bottom); \emph{Thick solid line (black):} dark matter;
  \emph{open circles:} galaxies in halos selected by $V_{max,now}$ at $z=0$; 
  \emph{thin solid line (blue):} predicted galaxy 3PCF using the dark matter 3PCF 
$Q_{dm}$ and eqn. (\ref{biasQ}) with 
best-fit bias parameters $c_1, c_2$ obtained from fitting the galaxy 3PCF    
at $r=9 h^{-1}$ Mpc (see \S \ref{sec:bias}); \emph{long dashed (red):} leading order 
non-linear perturbation theory prediction for dark matter reduced 3PCF.
Error bars calculated using jack-knife resampling method.}
\label{fig:alpha3PCF} 
\end{figure*}

Figures \ref{fig:equ3PCF} and \ref{fig:alpha3PCF} show the 3PCF 
for dark matter particles and
for galaxies from the L120 and L200 simulations. Here we plot
results for dark matter (thick solid line), galaxies in halos selected by
$V_{max,now}$ (thin solid line), and galaxies in halos selected by 
$V_{max,acc}$ (long-dashed line)
for $z=0$ and for halos selected by $V_{max,now}$ at $z=1$ (short-dashed
line) and $z=2$ (dotted line).  

Figure \ref{fig:equ3PCF} shows the reduced 3PCF for
equilateral triangles, $Q_{eq}(r)$, in real (left panel) 
and redshift space (right panel). In real space, 
the reduced 3PCF for the dark matter is only weakly scale 
dependent on small scales, decreases rapidly 
with increasing scale around $r \sim 3 h^{-1}$ Mpc, 
and falls off more slowly on larger scales. 
This behavior is broadly consistent with previous N-body results
\citep[e.g.][]{scoccimarro_etal:98}
and with expectations from leading order non-linear 
perturbation theory on the largest scales (shown as the thick red long-dashed
curve in Figure 1 left panel), 
with loop-corrected perturbation theory  
on intermediate scales where the rms perturbation amplitude $\delta(r)$
is of order unity (the transition to the strongly non-linear regime; 
adding more orders to the calculation would 
increase the agreement to the $N$-body dark matter 3PCF amplitude), 
and with quasi-stable hierarchical clustering on the smallest scales. 
Tests with the L200 box indicate that the downturn in $Q_{eq}$ for 
dark matter at scales larger than $r\sim 8 h^{-1}$ Mpc is likely due 
to finite volume effects.

At scales below $r \sim 10 h^{-1}$ Mpc, the dark matter $Q_{eq}$ is
larger than that for the galaxies; this behavior is 
broadly expected if galaxies are more strongly clustered than (positively 
biased with respect to) the mass, cf. eqn.(\ref{biasQ}). 
At higher redshift, evolution is seen in
$Q_{eq}(r)$ that is consistent with expectations 
from non-linear gravitational evolution: on the largest scales, 
the amplitude of $Q_{eq}$ is unchanging, as predicted from leading order 
perturbation theory, while the sharp break associated with the transition 
to the strongly non-linear regime moves to 
larger scales as the density perturbation amplitude increases with time. 

Comparing results for subhalos selected by $V_{max,now}$
and by $V_{max,acc}$, differences in the amplitude of $Q_{eq}$ appear on
small scales, $r \lesssim 3$ $h^{-1}$ Mpc. In halo-model language, 
on these scales the 3PCF is sensitive to the internal structure 
of halos, i.e., to the one- and two-halo terms, while the 
three-halo term dominates the 3PCF on larger scales 
\citep{wang_etal:04, takada_jain:03}. 

%

In Figure \ref{fig:alpha3PCF}, the top panels show how the reduced 3PCF
depends on triangle shape in real space.
In general, the 3PCF for elongated configurations is greater than for 
rectangular configurations.  This is a consequence of the fact that, in 
non-linear gravitational instability, velocity flows tend to occur along 
gradients of the density field \citep{bernardeau_etal:02}.
The 3PCF
is larger for dark matter than for galaxies for all shapes
and scales, although the difference is larger for elongated
configurations.  The difference in 3PCF amplitude between rectangular and
elongated configurations is larger on large scales, in broad agreement
with leading-order theoretical predictions \citep{bernardeau_etal:02}: on large scales, 
the strong shape dependence is determined by perturbative non-linear 
dynamics; on smaller scales, the shape dependence is washed out since 
the coherence between the velocity and density fields gives way to virialized 
motions. This scale dependence of the 3PCF shape is also reflected in the 
redshift evolution: in the L120 box (left panel), 
the galaxy 3PCFs at $z=1$ and 2 
(green-dashed and red-dotted curves) essentially retain 
the primordial shape dependence of leading-order non-linear perturbation theory, i.e., 
at those redshifts, these scales are still close to the quasi-linear regime.   
At $r=3$ $h^{-1}$Mpc, the largest evolution in $Q(\alpha)$ is found for
elongated configurations.

As was seen in Figure \ref{fig:equ3PCF}, the effect of changing halo 
selection from $V_{max,acc}$ to $V_{max,now}$ on the 3PCF shape appears 
only on small scales, $r \lesssim 1.5 h^{-1}$ Mpc, i.e., roughly 
within the scale of a typical cluster-mass host halo.

On the larger scales probed in the L200 box (right panel of 
Figure \ref{fig:alpha3PCF}), the galaxy reduced 3PCF (open circles) tracks 
the shape of the dark matter 3PCF fairly well. The difference between 
the galaxy and dark matter 3PCF amplitudes on these scales is reasonably 
well fit by a simple bias prescription: 
the thin blue curve is the biased 3PCF that results from fitting the galaxy 3PCF 
with 
eqn. (\ref{biasQ}); see \S \ref{sec:bias}. We also see that the jack-knife
errors increase on the largest scales, where the effects of the finite 
box size start to become evident. For comparison, 
the red long-dashed curve is the 3PCF of the dark matter
from leading-order non-linear perturbation theory \citep{bernardeau_etal:02,
  jing_borner:97}. On the largest scales, it is in reasonable agreement 
with the measured 3PCF for the dark matter.

\subsection{The 3PCF in redshift space}

Redshift distortions have been studied in depth (and are
a useful tool to constrain cosmological parameters) for the power
spectrum (e.g., \citealt{ 
bernardeau_etal:02, daangela_etal:05, tinker:06}) and for the bispectrum
\citep{scoccimarro_etal:99, verde_etal:02, sefusatti_etal:06}.  
Some comparisons have been made for the 3PCF as well 
\citep{matsubara_suto:94,takada_jain:03,wang_etal:04}. 
\cite{gaztanaga_scoccimarro:05} found that the redshift
distortions do not have a strong dependence on the 
cosmological parameters.

The right panel of Figure \ref{fig:equ3PCF} and the bottom panels 
of Figure \ref{fig:alpha3PCF} show the
3PCF in redshift space. The first feature that can be seen is a
dramatic decrease in the amplitude and in the scale and shape
dependence of $Q$ compared to the real-space measurements. 
For example, for equilateral triangles, the redshift space $Q_z(r)$ is reduced 
compared to the real space $Q(r)$ at
small scales and increased with respect to the real-space results on
larger scales. The overall effect is that $Q_z(r)$ is nearly 
scale-independent, i.e., the clustering appears more hierarchical in redshift space
\citep{suto_matsubara:94,matsubara_suto:94,scoccimarro_etal:99}. 
Moreover, in redshift space, the suppression of the galaxy 3PCF relative 
to that of the dark matter is apparent on all scales; it appears 
to be relatively independent of scale and configuration and is larger 
than the relative suppression in real space, consistent with 
earlier results \citep{wang_etal:04,gaztanaga_etal:05}.
It also appears that 
there is very little redshift evolution of the galaxy 3PCF in redshift space; 
the measurements at $z=1$ and $z=2$
are nearly indistinguishable from each other. With regard 
to halo selection, as in real space we find that $Q_{Vmax,now} <
Q_{Vmax,acc}$, but the differences between them are smaller than in 
real space.

Together, these results suggest that the shape and scale dependence of 
the reduced 3PCF in redshift space on the scales shown here are largely 
determined by redshift distortions, with non-linear gravitational evolution 
playing a subdominant role.

\section{Observing the 3PCF: Luminosity and color dependence}
\label{sec:lumcolor}

To investigate the dependence of the three-point clustering
on galaxy  luminosity and color  and to make direct  comparisons with
measurements from  recent redshift surveys, we calculate  the 3PCF for
galaxies with luminosity and color cuts similar to those that
have been applied to redshift survey data samples, with luminosity and
color information obtained as described in $\S  2$.  The luminosities
are assigned according to $V_{max,now}$  in the L120 and  L200 boxes, since
we hace those measurements for both boxes. As seen in the previous section,
 the use of $V_{max,now}$  is justified since the differences with respect
to $V_{max,acc}$ are only significant on small scales ($r \lsim 1.5$ $h^{-1}$Mpc)
in real space and are almost completely dissapear on redshift space.
We adopt the same binning scheme used in the previous section.

\begin{figure}
\plotone{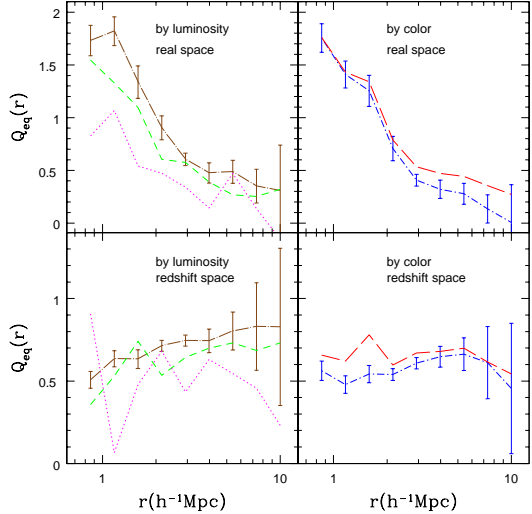}
\caption{The reduced 3PCF, $Q(r)$, for equilateral triangles as a
  function of galaxy luminosity and color in real and redshift
  space. \emph{Top left}: $Q_{eq}(r)$ in real space for galaxies divided
  into luminosity bins; \emph{long dash-dotted (cyan):} 
  $-19>\Mr>-20$;
  \emph{short-dashed (green):} 
  $-20>\Mr>-21$; \emph{dotted (magenta):} 
  $-21>\Mr>-22$. The brightest sample comes from the L200 box, the other two 
are from the L120 box. \emph{Bottom left:}
  Results in redshift space; line types correspond to the same galaxy
  samples as in the top left panel.  \emph{Top right:} $Q_{eq}(r)$ in real
  space for galaxies divided according to color, using the L120 box;
  \emph{long-dashed (red):} red galaxies ($g-r>0.7$); 
  \emph{short dash-dotted (blue):} blue
  galaxies ($g-r<0.7$). \emph{Bottom right} Results in
  redshift space; line types correspond to the same galaxy samples as
  in the top right panel. Error bars are calculated using jack-knife
respampling.}
\label{eq:LandC}
\end{figure}
%

\subsection{Luminosity dependence}

The left panels of Figure \ref{eq:LandC} show results for the reduced
3PCF for equilateral configuratons in two luminosity bins, in real 
and redshift space.  For equilateral triangles, there
is a small difference in 3PCF between the luminosity samples in real space: 
the fainter galaxies have larger $Q(r)$ than the brighter
ones, as expected in a simple linear bias model (eqn. \ref{biasQ}).
The redshift-space 3PCFs for these galaxies are roughly constant
with $r$, $Q_{z,gal}(r)\sim0.7$, in agreement with SDSS measurements
for these configurations (see Figures 7-9 in 
\citealt{kayo_etal:04}).  There is a very slight difference between
the reduced 3PCF amplitudes for different luminosity bins in redshift space, 
but it 
is not statistically significant for a dataset of this size.  This
result qualitatively agrees with the SDSS results of \cite{kayo_etal:04} 
who also
found almost no luminosity dependence of the reduced 3PCF on these scales.  
They found
a slightly higher amplitude for the reduced 3PCF for the $-19>\Mr>-20$ sample 
compared to that of 
brighter galaxies, but their results for the two luminosity samples  
were consistent within the error bars. The brightest galaxies (dotted
line) show significant fluctuations in both real and redshift space; 
we think this behavior is due to 
the small size of the box and the low density of objects.

\begin{figure}
\plotone{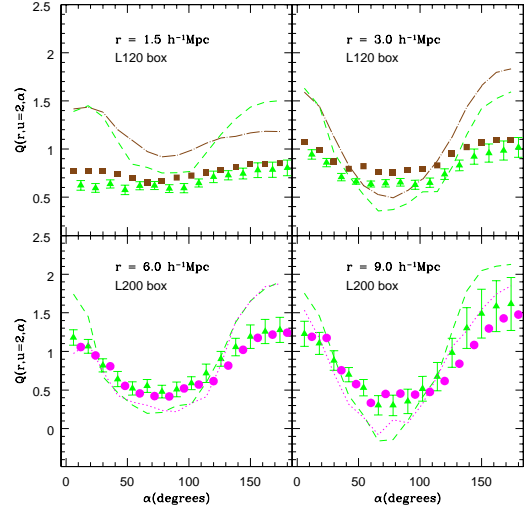}
\caption{$Q(\alpha)$ shape dependence for galaxies divided by luminosity, for 
  $r=$ 1.5, 3, 6, and 9 $h^{-1}$Mpc, with fixed $u=2$ in real (lines) 
  and redshift space (symbols). Top two plots are results for the L120 box: 
  \emph{filled squares and long dash-dotted (cyan):} $-19>\Mr>-20$;
  \emph{filled triangles and short-dashed (green):} $-20>\Mr>-21$;
  triangles are slightly shifted to the right for clarity.
  Bottom plots show results for the L200 box: \emph{filled triangles and short-dashed 
    (green):}  $-20>\Mr>-21$; \emph{filled circles and dotted 
    (magenta):} $-21>\Mr>-22$;
  circles are slightly shifted to the right for clarity. Error bars are
calculated using jack-knife resampling and are shown only for one of the 
samples.}
\label{fig:galqalpha_lum}
\end{figure}

\begin{figure}
\plotone{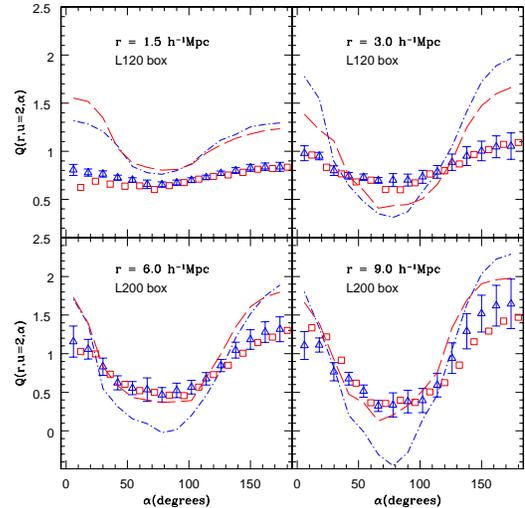}
\caption{
  $Q(\alpha)$ shape dependence for galaxies divided by color, 
  in the L120 box (top) and L200 box (bottom) in real (lines) and 
  redshift space (symbols).  \emph{Open squares and long-dashed line (red):}
  red ($g-r>0.7$) galaxies. \emph{open triangles and short dash-dotted 
(blue):} blue ($g-r<0.7$) galaxies; squares are slightly shifted to the
right for clarity. Error bars are
calculated using jack-knife resampling and are shown only for one of the 
samples.}
\label{fig:galqalpha_col}
\end{figure}

Figure \ref{fig:galqalpha_lum} shows the dependence of $Q(\alpha)$
on galaxy luminosity in real (curves) and redshift space (points). 
We use the same ordinate scale for all the plots
to emphasize where configuration effects are more important.  The top
panels show results for small scales, calculated with the L120
box.  On these scales, the reduced redshift-space 3PCF of the
brighter sample ($-20>\Mr>-21$, short dashed curve for real space, 
filled triangles for redshift space) is slightly lower 
than for the fainter sample ($-19>\Mr>-20$, long dashed-dotted
 curve for real space,
filled squares for redshift space) for all angles,
consistent with the results for $Q_z(r)$ using equilateral
triangles. 
The lower plots in Figure \ref{fig:galqalpha_lum} show the luminosity dependence 
on larger scales,
measured using the L200 box; note that the luminosity bins in the lower 
plots are $-20> \Mr > -21$ and $-21 >\Mr>-22$. 
The characteristic U-shape of $Q(\alpha)$ appears clearly in the
redshift-space measurements on scales larger than $r=6$ $h^{-1}$Mpc. 
The luminosity dependence of $Q$ on these scales appears non-existent  
in redshift space and only slight in real space.
There are hints that the reduced 3PCF for fainter galaxies may have slightly
higher amplitude and stronger shape dependence than for brighter galaxies,
but these trends are not statistically significant in the samples 
studied here. Rather, the strong luminosity dependence observed for the 
2PCF \citep{zehavi_etal:05} appears to be closely matched by a correponding 
dependence of the 3PCF, such that the reduced 3PCF $Q$ is roughly
independent of luminosity.  

\subsection{Color dependence}

The right panels of Figure \ref{eq:LandC} show the reduced 3PCF $Q(r)$ for equilateral
triangles separately for red and blue galaxies in real (top) and redshift space (bottom). 
Due to limited statistics, we use the full luminosity range in each color bin, 
as opposed to
\cite{kayo_etal:04}, who divided the color bins into luminosity subsamples as
well (see their Figure 9). Since the luminosity functions for red and blue 
galaxies differ, our red and blue samples have different characteristic 
luminosities --- the red sample is on average brighter. 
In both real and redshift space, the red galaxies appear to have a 
slightly higher reduced 3PCF amplitude. Comparing with the left 
panels, this difference is in the opposite sense from that expected 
from the fact that the red galaxies are brighter; put another way, if we were 
to compare red and blue samples of the {\it same} luminosity, the color difference 
in $Q$ would likely be larger than that seen here. On the other hand, we 
should not overinterpret these trends, since the differences are within 
the statistical errors. \cite{kayo_etal:04} found
a similarly weak dependence of $Q$ in redshift space on color.

Figure \ref{fig:galqalpha_col} shows the color dependence of
$Q(\alpha)$ in real (curves) and redshift space (points), for the same samples and 
configurations as in Figure
\ref{fig:galqalpha_lum}. In real space, on scales larger than about 3 $h^{-1}$ Mpc, 
$Q$ for red galaxies is larger for rectangular configurations and smaller 
for elongated configurations than for blue galaxies. This behavior is 
qualitatively consistent with a picture in which red galaxies preferentially 
occupy the inner regions of clusters, while blue galaxies tend to trace out 
more elliptical or filamentary structures. In redshift space, the trend with 
color is largely washed out. These redshift space results appear more 
consistent with the observations of the 2dFGRS (\citealt{gaztanaga_etal:05}),
where the color differences for $Q$ in redshift space are smaller than those seen in 
the SDSS by \cite{kayo_etal:04}. 

The fact that red galaxies have a larger two-point clustering amplitude 
than blue galaxies and that the reduced 3PCF for red galaxies is also 
larger than for blue galaxies (at least for rectangular configurations) 
suggests that red galaxies have a larger quadratic (non-linear) bias, 
Cf. eqn. \ref{biasQ}. This is qualitatively consistent with the observed 
morphology-density or color-density relation according to which red galaxies 
are preferentially found in dense regions, since the latter contribute 
more strongly to the higher-order correlations. 
The qualitative agreement between the two- and three-point 
observations and our method for assigning galaxy colors confirms that
the color of a galaxy, a consequence of many physical processes
occurring inside the galaxies, depends largely on the surrounding
environment.  However, we note that the color assignment for these
subhalos is the most uncertain part of the model; future work will be
required to determine how clustering statistics will change if more
sophisticated schemes are adopted.

%
\begin{figure}
\plotone{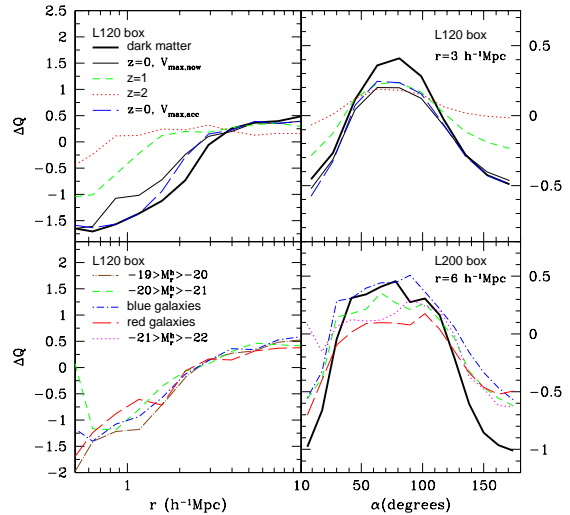}
\caption{$\Delta Q(r,u,\alpha) \equiv Q_z(r,u,\alpha)-Q_r(r,u,\alpha) $ 
as a function of galaxy type and redshift. 
\emph{Top:} $\Delta Q$ for dark matter at $z=0$ and for galaxy 
samples at different redshifts in the L120 box,
for equilateral triangles (\emph{top left}) and as a function of angle  
for triangles with $r=3$ $h^{-1}$Mpc and $u=2$ (\emph{top right}).
\emph{Bottom:} $\Delta Q$ as a function of galaxy type for equilateral
triangles in the L120 box (\emph{bottom left}) and as a function of angle for
$r=6$ $h^{-1}$Mpc and $u=2$ in the L200 box (\emph{bottom right}).}
\label{fig:deltaQ}
\end{figure}
%

\subsection{Redshift distortions: Galaxy Type and Evolution}
\label{subsec:redshift}

So far, we have explored some differences between measurements
in real and redshift space of the 3PCF for different simulated 
galaxy samples. Here we investigate in more detail whether the redshift
distortions of the 3PCF are universal or instead depend on the type 
of galaxy studied and to what extent they evolve with time.
We study the behavior of the quantity
\begin{equation}
\Delta Q (r,u,\alpha)\equiv Q_z(r,u,\alpha)-Q_r(r,u,\alpha),
\end{equation}
where $Q_r$, $Q_z$ represent the reduced 3PCF measured in real and redshift
space, respectively. We explore  $\Delta Q$ 
as a function of galaxy type (luminosity and color) and epoch. 
Figure \ref{fig:deltaQ} shows $\Delta Q$ for equilateral triangles 
as a function of scale and also the configuration dependence for 
triangles with fixed $r$, $u=2$, and
different opening angles $\alpha$. 

In general, the trends are similar to those seen above:
$Q_z<Q_r$ for small scales and for elongated 
triangle configurations, while the opposite behavior is seen for 
larger scales and for rectangular configurations. 
At $z=0$, for equilateral triangles $\Delta Q(r)$ appears to display 
a roughly universal scale dependence, independent of galaxy type, 
with $ \Delta Q(r) \simeq 0.67r^{-2}-2.6r^{-1}-0.02r+0.88$ over the 
range $0.5 \le r \le 5$ $h^{-1}$Mpc. On the other hand, for 
$r=6 h^{-1}$ Mpc and $u=2$, the shape dependence of $\Delta Q$ 
shows more dependence on galaxy type, with blue galaxies having 
larger values than red galaxies and bright galaxies smaller values 
than faint galaxies. Note that the redshift distortions of $Q$ 
appear insensitive to whether the subhalos are identified at the present 
or at the time they are first accreted onto host halos. 

We see clear evolution of $\Delta Q$ with redshift, as expected 
since redshift distortion effects should become more pronounced 
as perturbations become more non-linear. For
equilateral triangles, the upper left panel of Figure \ref{fig:deltaQ}
shows that the 
scale $r$ where $\Delta Q \sim 0$  
increases with time. The shape dependence of $\Delta Q$ at fixed 
scale also appears to increase with time.

\section{Observing the 3PCF: Comparison to SDSS Data and Binning Effects}
\label{sec:compare}

Below, we compare our results with 
measurements of the 3PCF from the SDSS by \cite{nichol_etal:06}.  The SDSS 
sample is magnitude limited, with
$m_r<17.5$, and has additional cuts in absolute magnitude, 
$-19<M_r<-22$, and in redshift, $0.05<z<0.15$.  
In order to compare with this data, 
we randomly resample the galaxies from the L120 
simulation (in particular, the $V_{max,noew}$ sample at $z=0$),
so that they have the same absolute magnitude distribution 
as the SDSS flux-limited sample used in \cite{nichol_etal:06}; 
we call these pseudo-flux-limited samples. This resampling technique does 
not properly model the distance-dependent selection function of the 
SDSS, but it should reproduce its clustering properties on average.
Moreover, since, as we have just shown, the luminosity
dependence of the reduced 3PCF is weak, we expect that this 
procedure should be sufficiently accurate for our purposes.

\begin{figure}
\plotone{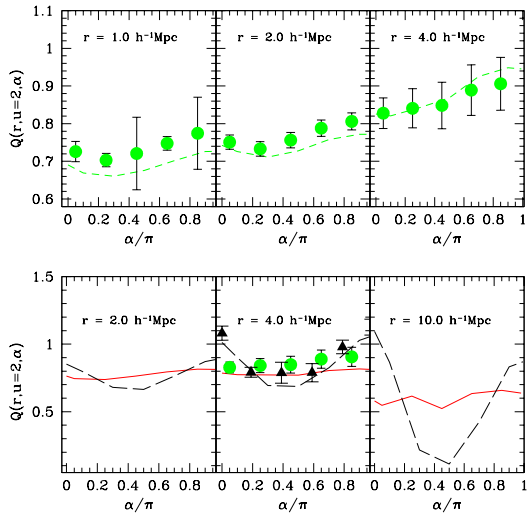}
\caption{\emph{Top:} $Q_z(r,u=2,\alpha)$ at $r=1.0$, 2.0 and 4.0 $h^{-1}$Mpc
  from SDSS observations (green points) from \cite{nichol_etal:06} and
from our pseudo-flux-limited-sample in the L120 box (long-dashed lines) with
the same binning scheme used in the mentioned paper.
\emph{Bottom:} 
Effect of binning in the 3PCF measurements.  $Q_z(r,u=2,\alpha)$ at $r=2.0$ and
4.0 $h^{-1}$Mpc in the L120 box and $r=10$ $h^{-1}$Mpc in the L200 box
 for the simulations in a volume-limited sample. \emph{solid (red):} results
using a wide binning scheme \citep{nichol_etal:06}; \emph{long-dashed (black):}
results with narrow binning. In the middle panel, we plot the results
using SDSS data with wide (green points) and narrow (triangles) binning.}
\label{fig:nichol}
\end{figure}

\cite{nichol_etal:06} measure the shape dependence of $Q(\alpha)$ for
four different length scales, using wide bins in $r,u$, and $\alpha$
(see their paper for more details): $\Delta r=1$ $h^{-1}$Mpc,
$\Delta u= 1$ and $\Delta \alpha=0.1$ rad.  In Figure \ref{fig:nichol}, 
we compare the SDSS results
(green points) to the calculation of $Q_z(r,u=2,\alpha)$ using our
model redshift-space, volume-limited sample 3PCF (solid lines) and our
pseudo-magnitude-limited sample (dotted line), each 
{\it with the same binning as the data}. 
We use the L120 box for small scales and
the L200 box for the $r=10$ $h^{-1}$Mpc measurements.
As the top panels of Figure \ref{fig:nichol} 
show, the model and data show good agreement 
within the model jack-knife error bars in amplitude as well as in
the shape of $Q(\alpha)$.

%

Over the range of scales shown here, both the SDSS data and the 
simulations show relatively little shape dependence for the reduced 
3PCF amplitude. This is in contrast to the
results of \S \ref{sec:lumcolor} above and to the 2dFGRS results of 
\cite{gaztanaga_etal:05} on similar scales, where 
a significantly stronger shape dependence of $Q(\alpha)$ is evident.
As noted by \citet{gaztanaga_scoccimarro:05} and \citet{kulkarni_etal:07}, 
the differences can be traced to the binning scheme: 
the relatively wide binning scheme used here results in smearing and 
therefore suppression of the U-shape of $Q$
over most scales of interest. This effect is not due 
primarily to the binning in $\alpha$: small bins
in both $r$ and $u$ are necessary to see the effects of 
shape-dependent clustering.

To illustrate these effects in more detail, in the lower panels of
Figure \ref{fig:nichol} we also 
show 
results for the same sample but with a narrower binning scheme, using 
$\Delta r = 0.1$ $h^{-1}$Mpc (ten times smaller), $\Delta u = 0.2$ (five
times smaller), and $\Delta \alpha = 0.05$ rad (two times smaller). 
With the narrower bins (shown by the dashed curves in Figure \ref{fig:nichol}), 
the shape-dependence of $Q$ is more pronounced 
for the simulation, especially on larger scales. For comparison, 
for $r=4$ $h^{-1}$Mpc (lower center panel), the solid triangles 
shows results for the SDSS flux-limited sample using a similar 
narrow binning scheme \cite{nichol_etal:06}, again showing good 
agreement between the model and the data.

\section{Galaxy bias and the 3PCF} 
\label{sec:bias}
As we have seen, the 3PCF predicted for galaxies differs systematically 
from that expected for dark matter. These differences reflect 
differences in the spatial distributions of these two populations;  
higher-order statistics can therefore provide important constraints upon the 
bias between galaxies and dark matter 
\citep{fry_gaztanaga:93,frieman_gaztanaga:94} 
and its dependence upon galaxy 
properties.  

On large scales, where the rms dark matter and galaxy overdensities are 
small compared to unity, 
it is common to adopt a deterministic, local bias model
\citep[e.g.][]{fry_gaztanaga:93}, 
\begin{equation}
\delta_{gal}=f(\delta_{dm})= b_1\delta_{dm}+\frac{b_2}{2}\delta_{dm}^2+...,
\label{delta-bias}
\end{equation}
where $\delta_{gal}$ and $\delta_{dm}$ are the local galaxy and 
dark matter overdensities smoothed over some scale $R$. 
We can use the simulations above to test how well this simple 
bias prescription characterizes the galaxy distribution and 
its clustering statistics.  

\begin{figure}
\plotone{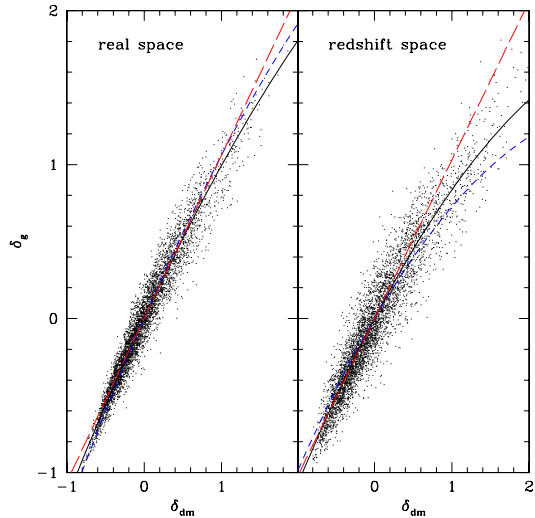}
\caption{Galaxy vs. dark matter overdensities for all galaxies in the 
L200 box measured in randomly placed spheres of radius 10$h^{-1}$ Mpc 
in real (left panel) and redshift space (right panel). Solid black 
curve denotes best fit using the quadratic bias relation of 
eqn. \ref{delta-bias}. Long-dashed red line indicates the best 
linear bias fit ($b_2=0$) 
to the 2PCF, i.e., estimating $b_1^2 = \xi_{gal}/\xi_{dm}$. Short-dashed 
blue curve indicates best fit of eqn. \ref{biasQ} to the 3PCF 
of the galaxies and dark matter.}
\label{fig:scatter}
\end{figure}
%

\begin{figure}
\plotone{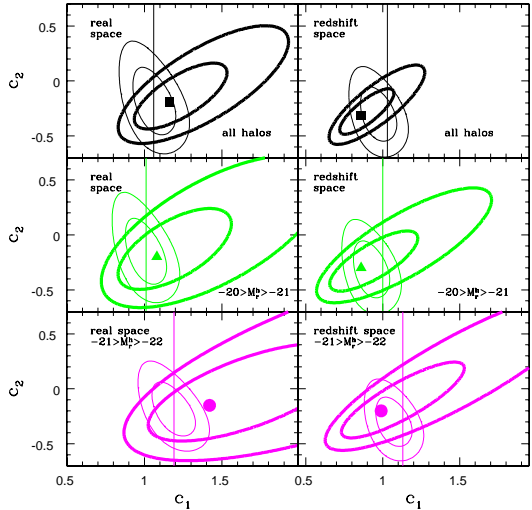}
\caption{Constraints on the model bias parameters $c_1$ and $c_2$ in 
real (left panels) and redshift space (right panels), 
measured in the L200 box for 
all galaxies (top), galaxies with $-20>\Mr>-21$ (middle), and
galaxies with $-21>\Mr>-22$ (bottom). 
Thick ellipses correspond to $1\sigma$ and $2\sigma$ 
constraints on parameters using the reduced
3PCF fit to equation \ref{biasQ}; symbols denote the 
minimum $\chi^2$ value. Thin ellipses come from fitting the 
smoothed halo
and dark matter overdensities (equation \ref{delta-bias}) shown in 
Figure \ref{fig:scatter}. The vertical
lines show the constraint on $c_1$ from comparing 
the dark matter and galaxy 2PCF amplitudes and fitting the 
ratio with a linear bias model, $c_1^2=\xi_{gal}/\xi_{dm}$.}
\label{fig:biasrz}
\end{figure}

Figure \ref{fig:scatter} shows the relation between 
$\delta_{gal}$ and $\delta_{dm}$ for all subhalos  in the L200 
box. The points show the overdensities for the galaxy and dark 
matter fields in randomly placed spheres of radius 10$h^{-1}$ Mpc 
in both real (left panel) and redshift space (right panel). The 
solid black curves show the best quadratic fits of the form 
in eqn. (\ref{delta-bias}). The 
quadratic local bias model appears to do a reasonable job in 
characterizing the mean relation. Nonetheless, there is  significant scatter, 
either due to stochastic bias or dependence of bias on other properties 
than $\delta_{dm}$, that is not captured by this simple bias model.
The errors on these fits, also  
extended to samples divided by galaxy luminosity, are shown 
in Fig. \ref{fig:biasrz} by the light solid contours. 

To leading order, this bias prescription leads to a relation 
between the galaxy and dark matter reduced 3PCF amplitudes of the form 
\citep{fry_gaztanaga:93,gaztanaga_scoccimarro:05}
\begin{equation}\label{biasQ}
Q_{gal} = \frac{1}{c_1}\left(Q_{dm}+c_2\right),
\end{equation}
where $c_1=b_1$ and $c_2=b_2/b_1$. Also to leading order, at 
low overdensities  
the relation between the galaxy and dark matter 2PCF amplitudes in this 
model 
is given just by the linear bias, $\xi_{gal} = b_1^2\xi_{dm}$.
We can test how well this bias prescription captures the 
clustering statistics 
by fitting these relations to the dark matter 
and galaxy 2- and 3-point correlation functions in the simulation in 
both real and redshift space and 
extracting the parameters $c_1$ and $c_2$. Since the relation 
(\ref{delta-bias}) holds for the density fields on some smoothing scale, we 
should fit the correlation functions for separations 
comparable to or slightly larger than this scale. 
For the 3PCF, we use triangles
with $u\equiv r_{23}/r_{12}=2$, 
$r=9$ $h^{-1}$Mpc, and weight equally all configurations with 
$0^0 < \alpha <180^0$ using the L200 box. To calculate the likelihood 
function, we 
use a method similar to that described in \citet{gaztanaga_scoccimarro:05},
which is based on an eigenmode analysis of the covariance matrix.
We use the jack-knife subsamples in real and redshift space 
to construct covariance matrices, and we use only the 
dominant eigemodes with values $>\sqrt{2/N}$ where $N=16$ is the
number of jack-knife subsamples; we found that adding further eigenmodes
just increases artificially the signal.

The best-fit parameter values 
from the 3PCF, substituted into Eqn. \ref{delta-bias}, are shown as 
the short-dashed blue curves in Figure \ref{fig:scatter}. Also, 
in the right panels of Figure \ref{fig:alpha3PCF} we show the 
predicted galaxy 3PCF (solid blue curves) using the measured 
dark matter $Q_{dm}$ and the best fitting $c_1, c_2$ parameters 
in Eqn. (\ref{biasQ}) from the fit at $r=9 h^{-1}$ Mpc. We see that
the agreement with the measured galaxy 3PCF is very good in both 
real and redshift space for triangles with $r=9 h^{-1}$ Mpc; 
for configurations with $r=6 h^{-1}$ Mpc, the agreement in real 
space is still quite good while in redshift space some deviations in 
the shape-dependence appear.

\begin{deluxetable*}{ l c c c c c}
\tablecaption{Best-fit bias parameters in the L200 box}
\tablehead{
\colhead{Subample} & \colhead{$c_1^Q$} & \colhead{$c_2^Q$}& \colhead{$c_1^{\delta}$} & \colhead{$c_2^{\delta}$} & \colhead{$c_1^{\xi}$}}
\startdata
All objects r-space & 1.16$^{+0.21}_{-0.15}$ & -0.19$^{+0.19}_{-0.16}$ &1.10$^{+0.05}_{-0.12}$&-0.20$^{+0.19}_{-0.22}$&1.06$\pm 0.09$\\
All objects z-space & 0.86$^{+0.11}_{-0.14}$ & -0.31$^{+0.15}_{-0.11}$ &0.97$^{+0.05}_{-0.07}$&-0.32$^{+0.12}_{-0.18}$&1.03$\pm 0.08$\\
$-20<\Mr<-21$ r-space & 1.08$^{+0.28}_{-0.18}$ &-0.20$^{+0.28}_{-0.23}$ &1.04$^{+0.07}_{-0.16}$&-0.08$^{+0.08}_{-0.19}$&1.01$\pm 0.09$\\
$-20<\Mr<-21$ z-space & 0.86$^{+0.21}_{-0.16}$ &-0.30$^{+0.20}_{-0.15}$ &0.97$^{+0.06}_{-0.12}$&-0.34$^{+0.22}_{-0.16}$&1.00$\pm 0.08$\\
$-21<\Mr<-22$ r-space & 1.42$^{+0.48}_{-0.29}$ &-0.15$^{+0.33}_{-0.35}$ &1.21$^{+0.06}_{-0.14}$&-0.36$^{+0.31}_{-0.05}$&1.19$\pm 0.10$\\
$-21<\Mr<-22$ z-space & 0.99$^{+0.15}_{-0.35}$ &-0.20$^{+0.21}_{-0.30}$ &1.01$^{+0.12}_{-0.05}$&-0.43$^{+0.33}_{-0.07}$&1.13$\pm 0.13$
\enddata
\label{tab:bias_parameters}
\end{deluxetable*}

In Figure \ref{fig:biasrz} we show the fits for $c_1$ and $c_2$ for
the different galaxy samples in the L200 box in both real and redshift
space: all galaxies (top panels), galaxies in the absolute 
range $-20>\Mr>-21$ (middle), and those in the range $-21>\Mr>-22$
(bottom).  The thick oval contours indicate the $1$
and $2\sigma$ confidence intervals for a $\Delta \chi^2$ distribution
with two free parameters, constrained using the 3PCF, via Eq.
(\ref{biasQ}), and the large points indicate the maxiumum likelihood 
values from the 3PCF. The thin solid 
contours show the constraints on $c_1$ and
$c_2$ using the fit of Eq. (\ref{delta-bias}) to the measurements 
in Figure \ref{fig:scatter}, and the vertical line in each panel shows the
estimate of $c_1$ from comparing the 2PCF amplitudes for galaxies 
and dark matter using pairs selected from the triplets used to 
measure the 3PCF. The best-fit values
are shown in Table \ref{tab:bias_parameters}: the first two
columns show the best-fit parameters $c_1^Q$, $c_2^Q$ using the reduced
3PCF (eq. \ref{biasQ}), the next two columns, $c_1^\delta$ and $c_2^\delta$, 
are calculated using the quadratic bias model (eq. \ref{delta-bias}) 
fit to the counts in cells, and in 
the last column $c_1^\xi$ is obtained from comparing the 2PCF of galaxies
and dark matter assuming a linear bias model, $c_1^2=\xi_{gal}/\xi_{dm}$.

In agreement with previous measurements in surveys and simulations,
the bias parameters obtained from the 3PCF are degenerate,
resulting in elongated contour ellipses; this could be mitigated to 
some extent by using a larger variety of triangle configurations.
In real space (left panels of Figure \ref{fig:biasrz}), 
we see that the three methods of extracting the bias parameters 
are in rough agreement: there is a preference for $c_1\sim 1$ and
negative $c_2$ as was found for the 2dFGRS 3PCF  measurements 
\citep{gaztanaga_etal:05}. 
Note that the 3PCF fit tends to overestimate $c_1$ and 
to slightly overestimate $c_2$ compared to the other two methods.
In redshift space (right panels), the opposite is true: 
the 3PCF constraint tends to underestimate $c_1$. The right 
panels of Figs. \ref{fig:scatter} and \ref{fig:biasrz} 
show that there is a larger 
discrepancy between the 3PCF fits and the counts-in-cells fit to 
Eqn. (\ref{delta-bias}) in redshift space, suggesting that the 
relation (\ref{biasQ}) may not be a good representation in redshift 
space.

%

While this comparison with the deterministic local bias model 
is suggestive, to test it more quantitatively one should use 
more triangle configurations and larger-volume catalogs to enable 
more precise calculation of the correlation matrices.

\section{Summary}
\label{sec:conclude}
We have studied the 3PCF of dark matter and galaxies in 
high-resolution dissipationless cosmological simulations.  The galaxy
model associates dark matter halos with galaxies by
matching the halo velocity function, including subhalos, with the
observed galaxy luminosity function by abundance, and has been shown
previously to provide an excellent match to observed two-point
statistics for galaxies \citep{conroy_etal:06}.  Our primary results
are as follows:

\begin{enumerate}

\item The reduced real-space 3PCF for both galaxies and dark matter has strong dependence on scale and shape. The shape dependence of the 3PCF strengthens with
  increasing scale, in agreement with previous simulation results for dark matter and with expectations from non-linear perturbation theory.
\item On small scales, or alternatively with increasing time, the 
  shape dependence of $Q$ washes out as virial motions within halos 
  replace coherent infall on larger scales.
\item Redshift-space distortions attenuate the shape and scale dependence 
of the reduced 3PCF and weaken the evolution with redshift. 
\item The reduced 3PCF shows only weak dependence on galaxy 
  luminosity and color; put another way, the scaling between the 3PCF 
amplitude and the 2PCF is predicted to be 
nearly independent of galaxy type in this model. The trend of $Q$ 
  with color is somewhat stronger than with luminosity: the reduced 
  3PCF is slightly enhanced for
  red galaxies over blue, especially for elongated configurations. 
\item Our model predictions are in excellent agreement with the shape and scale
  dependence of the galaxy 3PCF measured in the SDSS when
  the same binning scheme is used. Since the results are
  highly sensitive to the binning scheme, caution must be exercised in 
comparing theory and observations of the 3PCF. 
In combination with earlier results, this comparison
  indicates that a simple scheme in which galaxies and dark matter
  halos and subhalos are associated in a one-to-one fashion based on
  maximum circular velocity can provide a good match to a wide
  range of galaxy clustering statistics.
\item The effect on the 3PCF 
of changing the selection of subhalos ({\it i.e.,} of
  connecting galaxy 
luminosity to $V_{max,acc}$ instead of $V_{max,now}$) is evident
  on small scales ($r<2$ $h^{-1}$Mpc) and for elongated configurations,
  but is negligible on larger scales.  Future measurements of
  the 3PCF will help constrain different models for the
  association of galaxy luminosity and color with subhalo properties.
\item On scales of order 10$h^{-1}$ Mpc, a local, deterministic bias 
   scheme is in reasonable agreement with the galaxy and dark matter 
   distributions of the model. The bias parameters extracted from $Q_{gal}$ 
   are in reasonable agreement with the $\delta_{gal}$-$\delta_{dm}$ 
relation in real space, less so in redshift space. Nevertheless, 
  the redshift-space constraints on the bias parameters
  are in agreement with the 2dFGRS measurements of the 3PCF. 
\end{enumerate}

\acknowledgments 
We are indebted to Anatoly Klypin and Brandon Allgood
for running and making available the simulations used in this paper,
which were run on the Columbia machine at NASA Ames and on the Seaborg
machine at NERSC (Project PI: Joel Primack), to Andrey Kravtsov for
running some of the halo catalogs used in this study, and to Charlie
Conroy for providing us with measurements of $v_{\rm max, acc}$.  We
thank Enrique Gazta{\~n}aga for enlightening comments on measuring and
interpreting the 3PCF and comparing results of $Q(\alpha)$ with his
estimator; and Cameron McBride for discussions on the measurements of
the bias parameters.  We additionally thank Andrey Kravtsov, Issha
Kayo and Roman Scoccimarro for several useful discussions.  RHW was
partially supported by NASA through Hubble Fellowship grant
HST-HF-01168.01-A awarded by the Space Telescope Science Institute,
and also recieved support from the U.S. Department of Energy under
contract number DE-AC02-76SF00515.  This work was supported in part by
the Kavli Institute for Cosmological Physics through the grant NSF
PHY-0114422, and by the U.S. Department of Energy at Fermilab and at 
U. Chicago. FAM thanks the Fulbright Program and CONICYT-Chile for
additional support.

This study also made use of the SDSS DR3 Archive, for which funding
has been provided by the Alfred P. Sloan Foundation, the Participating
Institutions, the National Aeronautics and Space Administration, the
National Science Foundation, the U.S. Department of Energy, the
Japanese Monbukagakusho, and the Max Planck Society. The SDSS Web site
is http://www.sdss.org/.  The SDSS is managed by the Astrophysical
Research Consortium (ARC) for the Participating Institutions: the
University of Chicago, Fermilab, the Institute for Advanced Study, the
Japan Participation Group, the Johns Hopkins University, Los Alamos
National Laboratory, the Max-Planck-Institute for Astronomy (MPIA),
the Max-Planck-Institute for Astrophysics (MPA), New Mexico State
University, University of Pittsburgh, Princeton University, the United
States Naval Observatory, and the University of Washington.  We also
made extensive use of the NASA Astrophysics Data System and of the
{\tt astro-ph} preprint archive at {\tt arXiv.org}.

\bibliography{msv2}


\end{document}